# Integrated Longitudinal Speed Decision-Making and Energy Efficiency Control for Connected Electrified Vehicles

Teng Liu, Bo Wang, Dongpu Cao, Xiaolin Tang, Yalian Yang

*Abstract*—To improve the driving mobility and energy efficiency of connected autonomous electrified vehicles, this paper presents an integrated longitudinal speed decision-making and energy efficiency control strategy. The proposed approach is a hierarchical control architecture, which is assumed to consist of higher-level and lower-level controls. As the core of this study, model predictive control and reinforcement learning are combined to improve the powertrain mobility and fuel economy for a group of automated vehicles. The higher-level exploits the signal phase and timing and state information of connected autonomous vehicles via vehicle to infrastructure and vehicle to vehicle communication to reduce stopping at red lights. The higher-level outputs the optimal vehicle velocity using model predictive control technique and receives the power split control from the lower-level controller. These two levels communicate with each other via a controller area network in the real vehicle. The lower-level utilizes model-free reinforcement learning method to improve the fuel economy for each connected autonomous vehicle. Numerical tests illustrate that vehicle mobility can be noticeably improved (traveling time reduced by 30%) by reducing red light idling. The effectiveness and performance of the proposed method are validated via comparison analysis among different energy efficiency controls (fuel economy promoted by 13%).

The work was supported in part by Supported by the State Key Laboratory of Mechanical System and Vibration (Grant No. MSV202016). (Corresponding author: Bo Wang)

T. Liu is with College of Automotive Engineering, Chongqing University, Chongqing 400044, China, and also with Department of Mechanical and Mechatronics Engineering, University of Waterloo, N2L 3G1, Canada. (email: tengliu17@ gmail.com)

B. Wang is with School of Mathematics and Statistics, Beijing Key Laboratory on MCAACI, Beijing Institute of Technology, Beijing 100081, China. (email: wangbo89630@bit.edu.cn )

D. CAO is with Department of Mechanical and Mechatronics Engineering, University of Waterloo, Ontario N2L3G1, Canada. (dongpu.cao@uwaterloo.ca)

X. Tang and Y.Yang are with State Key Laboratory of Mechanical Transmissions, College of Automotive Engineering, Chongqing University, Chongqing, 400044, PR China. (email: tangxl0923@cqu.edu.cn, YYL@cqu.edu.cn)

## I. INTRODUCTION

In recent years, connected autonomous vehicle technology has aroused interest in the automobile companies as well as in the academic community [1]. Researchers are exploring ways to achieve mobility, safety and environmental impact goals via communication between vehicle and vehicle or infrastructure (V2V or V2I). Vehicle information (e.g. position, velocity, and acceleration) and traffic information (congestion and light timing) are able to be shared among the connected vehicles to reduce or eliminate crashes and enhance current operational practices of drivers [2].

To promote powertrain mobility and energy efficiency, reducing the red light idling and minimizing the braking times are the promising solutions for connected autonomous vehicles (CAVs) [3]. For example, Asadi et al. proposed a strategy using model predictive control (MPC) to decrease idle time at red lights and fuel consumption by signal phase and timing (SPAT) information [4]. A predictive optimal velocity planning algorithm was proposed in Ref. [5] by using the probabilistic traffic SPAT information. The relevant simulation results indicate that the presented method could increase energy efficiency via probabilistic timing and real-time phase data. Furthermore, Jin et al. proposed a power-based longitudinal control algorithm which can be applied in a connected eco-driving system [6]. The roadway grade and vehicle's brake specific fuel consumption are considered in this work when calculating an optimal speed profile with respect to energy savings and emissions reduction.

In current academic or industrial communities, CAVs are usually founded on or refitted from electrified vehicles, such as the hybrid electric vehicles (HEVs) and battery electric vehicles (BEVs) [7]. These vehicles have the potential to improve fuel economy and reduce pollutant emissions by embedding the electric storage systems into the powertrain [8]. Energy management is a significant technology for HEV and it means searching the optimal power distribution for a hybrid powertrain [9]. In general energy management problem, HEVs designers aim to optimize a pre-selected cost function (e.g. fuel consumption, harmful emissions, and running cost) by optimizing the power split controls of multiple energy storage sources while satisfying the driving power demand [10].

Many kinds of approaches have been utilized to formulate energy management strategies for HEVs, and the representative one is the optimization control theory-enabled method. For example, as the specific driving cycle is known a priori, dynamic programming (DP) algorithm [11] and equivalent consumption minimization strategy (ECMS) [12] have been developed to obtain the globally optimal power split decision for HEVs. The authors in [13] treated DP-based results as a benchmark to estimate the learning-based method in an energy management problem for a hybrid tracked vehicle. By computing the feasible range of optimal equivalent factor in ECMS, Ref. [14] presented an adaptive EMCS-enabled power split policy for a particular HEV configuration. Also, convex programming (CP) is another approach to derive the globally optimal power split control for HEVs [15]. The carbon reductions arising from renewable energy integration and the CP framework are quantified in [16], and the authors stated the proposed method could efficiently improve battery lifetime and reduce carbon dioxide emissions.

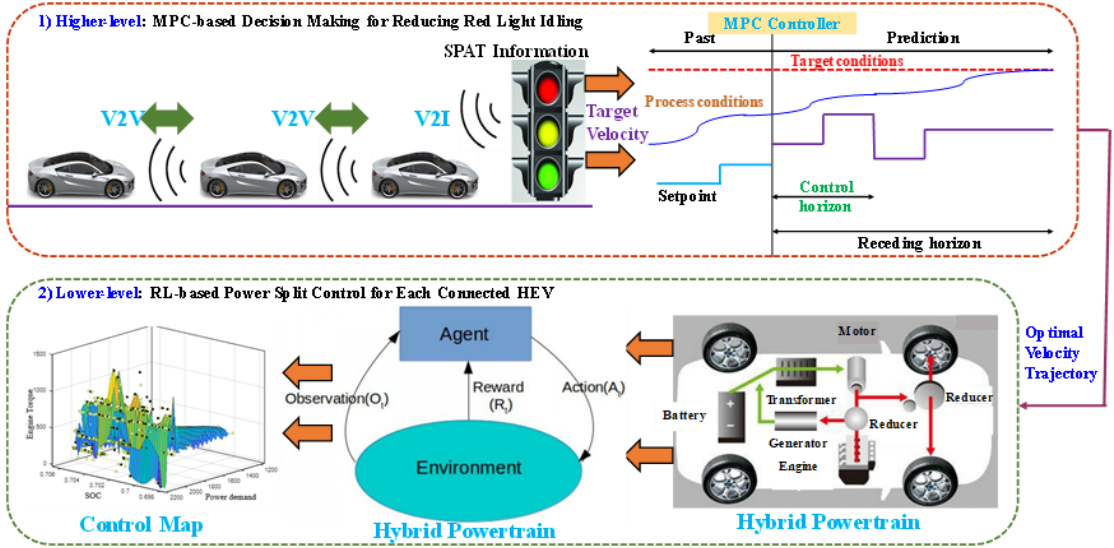

Fig. 1. Hierarchical architecture of the proposed decision-making and energy efficiency control system.

Furthermore, when the full driving cycle is not given prior, stochastic dynamic programming (SDP) [17], model predictive control (MPC) [18] and reinforcement learning (RL) [19] are studied to achieve real-time optimal controls. For example, Stellato combined approximate DP and direct MPC to handle integer optimal control problems over long prediction horizons [20]. It is able to reduce the computational burden and enable sampling times below 25 μs. Ref. [21] discussed the control performance of different RL algorithms, Q-learning and Sarsa, in an online energy management problem. The pros and cons of these algorithms are analyzed and compared in the same hybrid powertrain. However, most literature did not involve the energy efficiency and fuel economic controls of autonomous electrified vehicles in the connected environments.

Overall, an integrated combination of CAV and HEV attracted more and more attention in recent years. How to improve powertrain mobility and fuel economy of this synthetical architecture is an important and comprehensive research hotspot [22]. However, few literature has considered and discussed the automation function of CAV and energy utilization of HEV together [23]. In order to fill that literature gap, this paper aims to optimize an integrated objective for a combination of CAV and HEV, which consists of the powertrain mobility and fuel economy. The proposed system is a hierarchical control architecture, which is assumed to consist of higher-level and lower-level controls, see Fig. 1 as an illustration. The entire goal of this work is promoting driving mobility and reducing fuel consumption for each automated electrified vehicle in the connected environment (formulated in (2) and (8)). The presented system aims to achieve this objective in two steps, the vehicle speed is optimized by the higher-level control first and then the energy management policy is derived in the lower-level.

Especially, MPC is utilized to solve the longitudinal speed decision-making problem in higher-level and RL is employed to research the energy management problem in lower-level. The higher-level exploits the SPAT and state information provided by V2I and V2V communication to reduce stopping at red lights. By doing this, the waiting time at the signal lights can be availably shortened in order to promote efficiency and mobility. The higher-level outputs the optimal vehicle velocity for each vehicle and receives the power split control from the lower-level controller. The lower-level aims to improve fuel economy for each vehicle, which indicates it would manage the power distribution between different energy storage systems according to the obtained speed profile.

Three perspectives are contributed to this paper: (1) a novel hierarchical control architecture is presented to provide a holistic method for improving mobility and energy efficiency for connected autonomous electrified vehicles; (2) MPC approach is applied to make a decision that focuses on improving powertrain mobility and enhancing vehicle safety via decreasing red light idling; (3) model-free RL technique is applied to obtain optimal power split control between the engine and battery for each electrified vehicle. Numerical tests illustrate that vehicle mobility can be noticeably promoted by reducing red light idling. Comparison of different energy efficiency control strategies demonstrates the effectiveness and control performance of the presented method.

The rest sections of our paper are organized as follows. Section II describes the higher-level architecture of MPC framework for longitudinal speed decision-making in con-gested road conditions; Section III illustrates the formulation of a power split control problem in electrified powertrain, and the structure of the lower-level model-free RL algorithm is also introduced; In Section IV, the vehicle and experiment parameters are first elaborated, and experiment results of higher-level and lower-level performance comparison are presented; Section V summarizes the key takeaways.

## II. HIGHER-LEVEL PROBLEM FORMULATION AND MODEL PREDICTIVE CONTROLLER

In the higher-level, we focus on optimizing the traction or braking force to minimize the red light idling and energy consumption for traction of each connected autonomous vehicle. The position and velocity information of each vehicle can be obtained by V2V communication for their near neighborhood. The SPAT information of traffic lights is sent to each vehicle by roadside units through the wireless network. The proposed controller consists of two parts: First, a target velocity is calculated to avoid stopping at red lights, and second, MPC strategy is formulated to minimize the pre-defined cost function.

### A. Higher-Level Problem Formulation

For longitudinal control, the dynamics of the vehicle $i$ is described as [24]:

$$\begin{cases} \dot{x}_i = f_i(x_i, u_i) \\ f_i(x_i, u_i) = \begin{bmatrix} v_i \\ -\dfrac{1}{2M_i} C_d \rho_a A_i v_i^2 - \mu g - g\sin\theta + u_i \end{bmatrix} \end{cases} \quad (1)$$

where $x_i = [s_i, v_i]$, $s_i$ and $v_i$ denote the position and velocity of vehicle $i$. The control action $u_i$ is the traction or braking force per unit mass. The symbols $M_i$, $C_d$, $\rho_a$, $A_i$, $\mu$ and $\theta$ denote vehicle mass, drag coefficient, air density, vehicle frontal area, rolling friction coefficient, and road gradient, respectively.

In the higher-level, the optimization objective can be computed by dividing the total distance from fuel consumption, which is denoted by $\sum \dot{m}_f^i(t) \Delta t / s_i^{t_f}$. Here $\dot{m}_f^i(t)$ is the fuel consumption rate, $\Delta t$ is the time step, and $s_i^{t_f}$ is the distance traveled by vehicle $i$ during the time interval $[0, t_f]$. The following expression describes the optimal control problem in higher-level [22]:

$$\begin{cases} \arg\min_{u_i(t)} \sum_{i=1}^{n} \dfrac{1}{s_i^{t_f}} \sum_{t=0}^{t_f} \dot{m}_f^i(t)\Delta t \\ \dot{m}_f^i(t) = \dfrac{1}{\eta_{tr}^i H_{LHV}^i} P_w^i(t) \\ P_w^i = \alpha_i v_i^3 + M_i g v_i (\mu + \theta) + \beta_1^i [v_i M_i u_i] + \beta_2^i [-\eta_{rec}^i v_i M_i u_i] \end{cases} \quad (2)$$

$$s.t. \quad \alpha_i = \dfrac{1}{2} C_d \rho_a A_i, \; \beta_1^i = \begin{cases} 0 \; if \; u_i \leq 0 \\ 1 \; otherwise \end{cases}, \; \beta_2^i = \begin{cases} 0 \; if \; u_i > 0 \\ 1 \; otherwise \end{cases}$$

$$v_{min} \leq v_i \leq v_{max}, \; u_{min}^i \leq u_i \leq u_{max}^i$$

where $\eta_{tr}^i$ is the transmission efficiency connecting the fuel tank and powertrain, $H_{LHV}^i$ represents the fuel lower heating value for vehicle $i$, $P_w^i$ is the traction power depending on the traction force, and $\eta_{rec}^i$ is the recuperation efficiency acquired from the lower-level controller [22]. $v_{min}$, $v_{max}$, $u_{min}^i$ and $u_{max}^i$ depict the upper and lower threshold values for the allowable speed and possible acceleration, respectively. Hence, the higher-level controller generates the optimal velocity profile for each connected vehicle and inputs them into the lower-level for energy efficiency control.

### B. Target Velocity Computation

The upcoming SPAT information is obtained a priori via the V2I communication for each connected vehicle. Based on this, the target velocity of each vehicle can be decided wisely to avoid idling at the red light. The target velocity at time instant $k$ for the vehicle $i$ is defined as following [23]:

$$v_{tar}^i(k) = \begin{cases} \dfrac{d_{ia}(k)}{K_w t_{cy} - t_g - k} & if \; Lig = red \\ v_{max} & if \; Lig = green \; and \; \dfrac{d_{ia}(k)}{K_w t_{cy} - k} \leq v_{max} \\ \dfrac{d_{ia}(k)}{K_w t_{cy} + t_r - k} & if \; Lig = green \; and \; otherwise \end{cases} \quad (3)$$

$$Lig = \begin{cases} red & if \; 0 \leq \mathrm{mod}(k/t_{cy}) \leq t_r \\ green & if \; t_r < \mathrm{mod}(k/t_{cy}) < t_{cy} \end{cases}$$

$$s.t. \quad t_{cy} = t_r + t_g, \quad v_{min} \leq v_{tar}^i(k) \leq v_{max}, \quad K_w > k/t_{cy}$$

where $d_{ia}(k)$ is the distance between traffic signal and the vehicle location ($s_i(k)$), $t_r$, $t_g$ and $t_{cy}$ denote the red, green light and total cycle durations, respectively. $K_w$ represents the cycle number for the traffic light, and $mod$ denotes a function to reserve the remainder for division $k$ by $t_{cy}$.

Eq. (3) indicates that when the equation $d_{ia}(k) / (K_w t_{cy} - k) \leq v_{max}$ is satisfied and the status of the traffic light is green, the target velocity is selected as the maximum allowable speed. Violation of this condition implies that the vehicle cannot pass through the current green light cycle although using the maximum speed. Thus, the vehicle needs to slow down and wait for the next green light window. Sometimes, the vehicle has to stop as there is no feasible velocity in the consecutive green light windows [23]. Overall, target velocity means the available speed to help each vehicle pass through the traffic signal at different green light windows.

Assuming the vehicle $i$ could go through the traffic signal at one of the green light windows, and thus there is a speed range for the target velocity because the green light window is a time interval. Assume $v_l^i(k)$ denotes the lower threshold velocity and $v_u^i(k)$ depicts the upper value, they are computed as [22]:

$$v_l^i(k) = \begin{cases} \dfrac{d_{ia}(k)}{K_w t_{cy} - k} & \text{if } Lig = red \\ \dfrac{d_{ia}(k)}{K_w t_{cy} - k} & \text{if } Lig = green \text{ and } \dfrac{d_{ia}(k)}{K_w t_{cy} - k} \leq v_{max} \\ \dfrac{d_{ia}(k)}{(K_w + 1) t_{cy} - k} & \text{if } Lig = green \text{ and otherwise} \end{cases} \quad (4)$$

$$v_u^i(k) = v_{tar}^i(k)$$

$$s.t. \quad v_{min} \leq v_l^i(k) \leq v_{tar}^i(k)$$

Fig. 2 depicts the schematic of the target velocity and velocity range calculation process [22]. $t_1$ and $t_2$ describe the time interval between the current time instant and the next green instant or the next red instant, respectively. Note that, the upper value of the speed range is the target velocity.

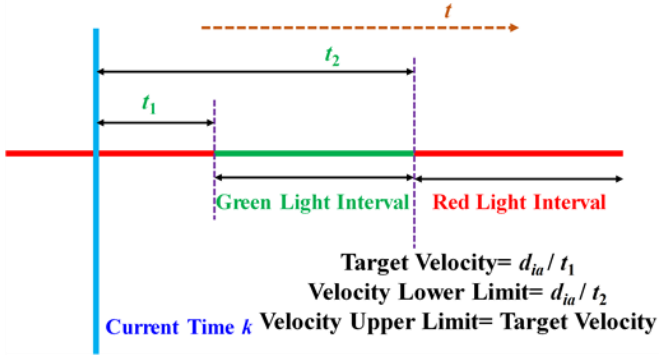

Fig. 2. Diagram for velocity range and target velocity computation

Based on the speed range, the limits on the control action can be calculated as follow:

$$v_l^i(k) \leq v_i(t) \leq v_u^i(k)$$
$$\Rightarrow v_l^i(k) \leq v_i(t-1) + \hat{a} \Delta t \leq v_u^i(k)$$
$$\Rightarrow \frac{v_l^i(k) - v_i(t-1)}{\Delta t} \leq \hat{a} \leq \frac{v_u^i(k) - v_i(t-1)}{\Delta t} \quad (5)$$

$$\text{where} \quad \hat{a} = u_i(t) + \hat{p}_i$$

$$\hat{p}_i = -\frac{1}{2 M_i} C_d \rho_a A_i v_i^2 - \mu g - g \sin \theta$$

Thus, the control action range (traction or braking force) is evaluated as:

$$\frac{v_l^i(k) - v_i(t-1)}{\Delta t} - \hat{p}_i \leq u_i(t) \leq \frac{v_u^i(k) - v_i(t-1)}{\Delta t} - \hat{p}_i \quad (6)$$

$$s.t. \quad u_{min}^i \leq u_i(t) \leq u_{max}^i$$

This limit implies that the vehicle velocity will fall into the velocity range as in Eq. (4), if the control action is bounded as in Eq. (6). Therefore, the vehicle can avoid stopping at a red light.

*C. Model Predictive Control*

In the previous section, we have discussed the optimization objective in the higher-level problem and the definition of the target velocity based on the communication between each vehicle and traffic lights via V2I information exchange. Note that, the optimal velocity trajectory for energy efficiency control and the target velocity in the higher-level may be different because the latter does not consider energy efficiency. Hence, we compute the optimum velocity profile by considering the extra position information of the front vehicle $j$ over a finite future time horizon $T$ via V2V information exchange. The discussed driving scenario is depicted in Fig. 3, wherein the driving mobility of each vehicle can be interpreted as the traveling time for the same scenario and the energy efficiency could be quantified as the fuel consumption of each HEV [25, 26].

As the position information of the front vehicle $j$ and the target velocity of the current vehicle $i$ are available, the MPC approach is used to generate the best velocity profile. It can be formulated as a nonlinear optimization problem at time instant $k$ as follow [27]:

$$J_i(k) = \arg\min_{u_i(t)} \left[ \sum_{t=k}^{k+T-1} \left[ \varphi_1 \frac{\dot{m}_f^i(t) \Delta t}{s_i^{k+T-1} - s_i^k} + \varphi_2 X_{ij}^2 \right. \right.$$
$$\left. \left. + \varphi_3 (v_i(t) - v_{tar}^i(k))^2 + \varphi_4 (u_i(t))^2 \right] \right] \quad (7)$$

$$X_{ij} = S_0 + t_{hw}(v_i(t) - v_j(t)) + (s_i^t - s_j^t)$$

$$s.t. \quad v_l^i(k) \leq v_i(t) \leq v_u^i(k), \hat{u}_{min}^i \leq u_i(t) \leq \hat{u}_{max}^i, \forall t$$

where $S_0$ and $t_{hw}$ represent the predefined critical distance and headway time, and the two inequality constraints are derived from Eqs. (4) and (6) respectively. In the cost function, the first term depicts the division of total fuel consumption by travel distance, the second items denote the desirable distance deviation between the vehicles $i$ and $j$, the third term restrains the velocity of vehicle $i$ close to the computed target velocity, and the last term aims to optimize control action [22].

The optimal energy efficiency velocity profile highly depends on the selection of the weights $\varphi_1$ to $\varphi_4$. The weights $\varphi_1$ and $\varphi_3$ are defined as a function of $(v_u^i(k) - v_l^i(k))$ to ensure when the deviation is large, more emphasis is focused on energy efficiency rather than velocity tracking ($\varphi_1$ high and $\varphi_3$ low), and as the deviation is small, more attention is paid to the target velocity tracking ($\varphi_1$ low and $\varphi_3$ high). The weight $\varphi_2$ is relevant with the distance deviation ($s_j^t - s_i^t$), and it increases as the deviation is small and it decreases as the deviation is large, finally the weight $\varphi_4$ is a constant [23]. In conclusion, the speed range $[v_l^i(k), v_u^i(k)]$ serves two goals, first it helps choose some weights in the cost function, and second, it forces the constraints on the velocity deviation to avoid red light idling.

Three key processes are involved in MPC algorithm: 1) search the optimization controls to minimize the pre-defined cost function during a prediction horizon; 2) obtain the first portion of the optimization controls and apply it into the physical plant; 3) moves forward the entire prediction horizon and repeat the first step [28]. Here we apply sequential quadratic programming to solve the nonlinear optimization problem (see Eq. (7)). Especially, the optimal control actions are computed using the command *fmincon* in Matlab/Simulink and the optimal solution for time $k$ is applied to formulate the initialization of the decision variables at time instant $k+1$. Overall, the MPC-based optimal velocity trajectory calculation may take about one minute in higher-level.

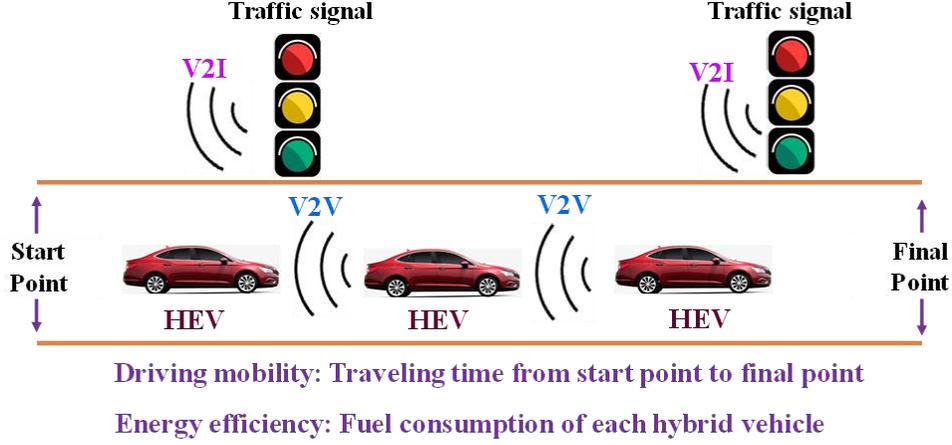

Fig. 3. Physical system of the studied connected HEVs.

## III. LOWER-LEVEL PROBLEM FORMULATION AND REINFORCEMENT LEARNING

In the lower-level, we aim to improve the energy efficiency of each connected autonomous electrified vehicle by deriving the optimal power split control between multiple energy sources. The sole state variable is the state of charge (SOC) in the battery and control action is the engine throttle percentage $th(t)$. After the higher-level gave the optimal speed profile, the relevant power demand can be calculated via the powertrain modeling. Then, the transition probability matrix (TPM) of the power demand can be calculated by the maximum likelihood estimate. Finally, the Q-learning algorithm in the RL framework is employed to evaluate the optimal control problem in the lower-level.

### A. Lower-level problem

In the minimization problem of the energy efficiency for a parallel electrified vehicle, the optimization objective is the sum of the charge sustenance and fuel consumption. The problem can be mathematically established as follow [20, 21]:

$$J = \arg\min_{th(t)} \sum_{t=0}^{t_f} [\dot{m}_f^i(th(t), SOC(t)) + \chi(SOC(t) - SOC(0))^2]\Delta t$$

$$s.t.\ SOC_{\min} \le SOC(t) \le SOC_{\max},\ th_{\min} \le th(t) \le th_{\max}$$
$$T_{en,\min} \le T_{en}(t) \le T_{en,\max},\ n_{en,\min} \le n_{en}(t) \le n_{en,\max} \quad (8)$$
$$P_{b,\min} \le P_b(t) \le P_{b,\max},\ I_{b,\min} \le I_b(t) \le I_{b,\max}$$

where $\chi$ is a positive weighting factor (equal to 1000 in this work) to guarantee the initial and final SOC values stay the same. $SOC_{\min}$, $SOC_{\max}$, $th_{\min}$, and $th_{\max}$ are the boundary values of the SOC and throttle percentage. $T_{en}$ and $n_{en}$ are the engine torque and speed, $n_{en,\min}$, $n_{en,\max}$, $T_{en,\min}$, and $T_{en,\max}$ are the permitted lower and upper bounds. $P_b$ and $I_b$ are the battery power and current, $P_{b,\min}$, $P_{b,\max}$, $I_{b,\min}$, $I_{b,\max}$, are the threshold of the admissible sets for them.

The output torque and speed of engine are decided by the throttle percentage at each time step. The fuel consumption rate in the engine and the engine output power are defined as:

$$\begin{cases} \dot{m}_f^i(th(t)) = f_{en}(T_{en}^i(t), n_{en}^i(t)) \\ P_{en}^i(t) = T_{en}^i(t) \times n_{en}^i(t) \end{cases} \quad (9)$$

Modeling the battery with the equivalent circuit model [29], the state equation of the SOC is given by:

$$\frac{dSOC}{dt} = -\frac{I_b(t)}{Q_b} = \frac{(V_{oc} - \sqrt{V_{oc}^2 - 4r_{in}P_b(t)})}{2Q_b r_{in}} \quad (10)$$

where $Q_b$ denotes the battery rated capacity. $V_{oc}$ represents the open circuit voltage and $r_{in}$ describes the battery internal resistance. $V_{oc}$ and $r_{in}$ are relevant to the SOC.

### B. Transition Probability Matrix

As the velocity profile is known in advance from the higher-level controller, the power demand to impel each connected autonomous electrified vehicle is computed as follow:

$$P_{dem}^i = \delta M_i a_i v_i + \frac{C_d A_i}{21.15} v_i^3 + M_i g \mu v_i \quad (11)$$

where $\delta$ is the mass factor, $a_i$ is the $i$-th vehicle acceleration.

For the purpose of keeping power balance in the powertrain, the power demand is satisfied by both of the engine and battery:

$$P_{dem}^i = (P_{en}^i + P_b^i \eta_m^i)\eta_{tr}^i \quad (12)$$

where $\eta_m^i$ is the traction motor efficiency for vehicle $i$. As the engine output power can be evaluated by Eq. (9), and then the battery power can be estimated from Eq. (12). The transition probability of the power demand is depicted via maximum likelihood estimator as:

$$\begin{cases} p_{nk,m} = P(P_{dem,m}^i | P_{dem,n}^i, v_k^i) = \frac{N_{nk,m}^i}{N_{nk}^i} \\ N_{nk}^i = \sum_{m=1}^M N_{nk,m}^i\ \ m,n = 1,2,...M \end{cases} \quad (13)$$

where $N_{nk,m}^i$ denotes the number of times for the transition from $P^i_{dem,n}$ to $P^i_{dem,m}$ at vehicle speed $v^i_k$, $N^i_{nk}$ is the total transition counts initiated from $P^i_{dem,n}$ at vehicle speed $v^i_k$, $k$ is the discrete time step, and $M$ is the amount of discrete power demand index. Fig. 4 describes the calculational schematic for the TPM of the power demand.

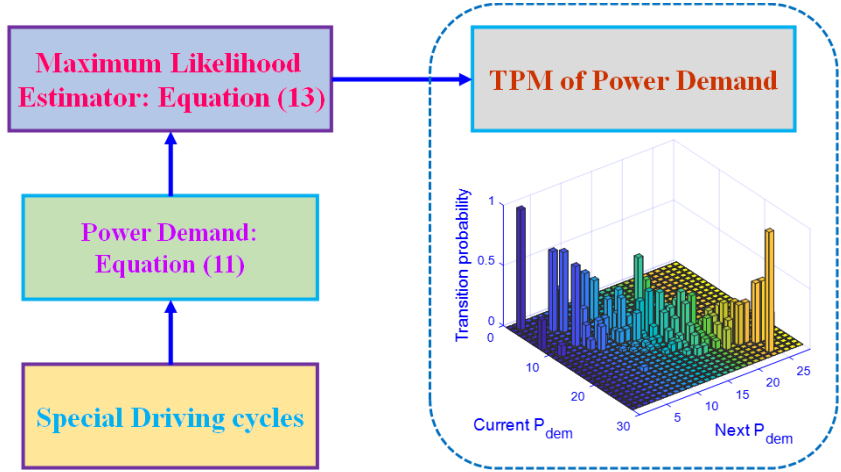

Fig. 4. Schematic for TPM of the power demand calculation.

### C. Reinforcement Learning Control

Based on the previous TPMs of power demand and vehicle modeling, RL technique is employed for energy efficiency control derivation. Especially, the Q-learning algorithm is implemented in Matlab with the Markov decision process (MDP) toolbox presented in [30], and the computer has an Intel quad-core CPU of 2.70 GHz a memory of 3.8 GB.

In the RL framework, a learning agent interacts with a stochastic environment. Five key variables are exploited to model the interaction, wherein $S$ is the state variables set, $A$ is the control actions set, $P$ denotes the TPM for power request, $r(s, a) \in R$ represents the reward function and $\tau \in (0, 1)$ means a discount factor.

The sequence of control actions constitutes the control policy $\pi$. The optimal value function can be computed as follows [31]:

$$V^*(s) = \min_{\pi} E(\sum_{t=0}^{t_f} \tau^t r(s,a))$$
$$= \min_a (r(s,a) + \tau \sum_{s' \in S} p_{sa,s'} V^*(s')) \quad \forall s \in S \quad (14)$$

where $p_{sa,s'}$ denotes the transition probability. As the optimal value function is decided at each state variable, the relevant optimal control policy is determined as:

$$\pi^*(s) = \arg\min_a (r(s,a) + \tau \sum_{s' \in S} p_{sa,s'} V^*(s')) \quad (15)$$

The computation process of optimal control action can be interpreted as the matching design of state variable and control action. Hence, the $(s, a)$ is called state-action pair and the related action-value function $Q(s, a)$ is given as follows:

$$\begin{cases} Q(s,a) = r(s,a) + \tau \sum_{s' \in S} p_{s'a,s} Q(s',a') \\ Q^*(s,a) = r(s,a) + \tau \sum_{s' \in S} p_{s'a,s} \min_{a'} Q^*(s',a') \end{cases} \quad (16)$$

Note that, variable $V^*(s)$ means the optimal value function that taking an optimal action $a$ at state $s$. Thus, $V^*(s) = Q^*(s, a)$ and $\pi^*(s) = \arg\min_a Q^*(s, a)$. Finally, the updated criterion for the optimal action-value function of the Q-learning algorithm is indicated by:

$$Q(s,a) \leftarrow Q(s,a) + \gamma(r(s,a) + \tau \min_{a'} Q(s',a') - Q(s,a)) \quad (17)$$

where $\gamma \in [0, 1]$ is a decaying factor of the Q-learning algorithm.

TABLE I
ITERATION PROCESS OF RL-BASED CONTROL

| Method: Q-learning Algorithm |
|---|
| 1. Give a value for $N_K$, $Q(s, a)$ and $s$ |
| 2. Repeat step by step ($k$=1, 2, 3…) |
| 3. Using $\varepsilon$-greedy policy to select control action $a$ |
| 4. Compute $r$, $s'$ based on $s$ and $a$ |
| 5. Compare and decide $a^*$=arg min$_a$ $Q(s', a)$ |
| 6. $Q(s, a) \leftarrow Q(s, a) + \gamma(r(s, a) + \tau$min$_{a'}$ $Q(s', a')$-$Q(s, a)$) |
| 7. Move to next step, $s \leftarrow s'$ |
| 8. finish when $s$ is the terminal |

To solve the optimal energy efficiency control problem, the state variables and control actions are discretized as follow: $S=\{(SOC(t))|0.4 \leq SOC(t) \leq 0.8\}$ and $A=\{th(t)| 0 \leq th(t) \leq 1\}$. The set of reward function is $R=\{\dot{m}_f(s, a) + \chi (SOC(t)-SOC(0))^2\}$. The initial and final values of SOC are selected to avoid overcharge and over-discharge in the battery. Table I shows the calculation process of optimal control actions in Q-learning algorithm. First, the matrix of action-value function $Q(s, a)$ and the state variable $s$ are initialized. Then, in each episode, the control action is decided by $\varepsilon$-greedy policy at each step, and the action value function is updated by (17). Finally, one episode ends when the state variable is the terminal, and the whole process ends when the number of episodes reaches the setting value. The obtained matrix $Q(s, a)$ can be used to derive the control action in particular driving situation.

In this paper, the discount factor $\tau$ is taken as 0.96 to balance the importance of current and future rewards, the decaying factor $\gamma$ is equal to $1/\sqrt{k+2}$ to accelerate the convergence rate and the value of $\varepsilon$ is $0.2*0.99^k$ ($\varepsilon$ decreases with time steps) to select the control action at each step. The iterative times $N_K$ is 10000, and the sample time $\triangle t$ is 0.5 second. The effectiveness of the MPC and RL approaches for the higher-level, and lower-level problems are validated in the next section.

### D. Parameters for SPAT and Powertrain

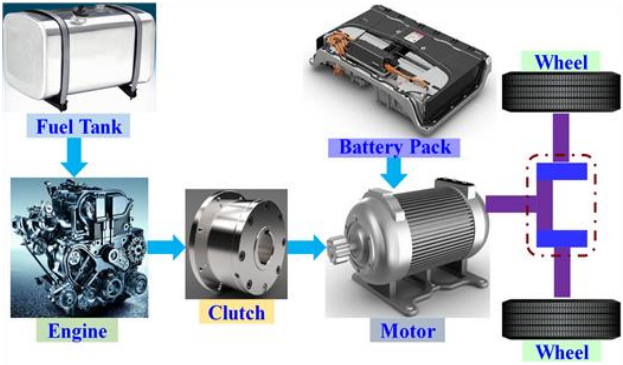

Fig. 5. Configuration of each connected electrified vehicle.

A signal lane road with eight vehicles constructs the simulation scenario, wherein the traffic lights are settled at every 500 m. Each parallel electrified vehicle has the same configuration and powertrain parameters, see Fig. 5 as an illustration. The primary parameters of them are listed in Table 2. The rated power for the diesel engine is 100 kW, and the maximum torque of diesel engine is 200 Nm as the rotational speed change from 110 rpm to 390 rpm. The maximum power, maximum torque and maximum rotational speed of the traction motor are 78 kW, 130 Nm, and 600 rpm, respectively. The battery pack is 6 Ah capacity with a nominal voltage of 250 V.

TABLE II
CHARACTERISTICS OF THE PARALLEL ELECTRIFIED POWERTRAIN

| Variables | Names | Numbers |
|---|---|---|
| $M_i$ | Vehicle weight | 1000 kg |
| $A_i$ | Fronted area | 2.25 m$^2$ |
| $C_d$ | Aerodynamic coefficient | 0.3 |
| $\eta^i_{tr}$ | Efficiency of Transmission axle | 0.9 |
| $\eta^i_m$ | Efficiency of Traction motor | 0.95 |
| $\mu$ | Coefficient of Rolling resistance | 0.008 |
| $\theta$ | Road gradient | 0 |
| $\rho_a$ | Air density | 1.205kg/m$^3$ |

For the SPAT parameters, the time intervals for the red and green light are 30 and 15 seconds ($t_r$=30 and $t_g$=15), respectively. The minimum and maximum allowable speeds for each connected vehicle are unified as 0 m/s and 20 m/s. Every vehicle's initial position is given by random, and the initial relative distance between any two vehicles is also from 10 m to 20 m by random. The initial velocity of each connected autonomous electrified vehicle is also stochastic and selected from 10 m/s to 15 m/s.

## IV. RESULTS AND DISCUSSION

The proposed MPC-enabled and RL-based longitudinal speed decision-making and energy efficiency control strategy for connected autonomous electrified vehicles is evaluated in this section. First, the performance of the MPC-enabled higher-level controller is examined by comparing with a baseline method. Then, the presented hierarchical system is compared with a benchmark system to assess its availability. The benchmark system employs MPC method in both higher and lower levels. Simulation results demonstrate that the proposed longitudinal speed decision-making and energy efficiency control strategy can highly reduce stopping at red lights and improve fuel efficiency.

### A. Higher-Level Controller Evaluation

In this section, the MPC-based longitudinal speed decision-making control is compared with a baseline method to verify its effectiveness. Here, the baseline method means the connected vehicle apply V2I communication to calculate the target velocity in Eq. (3), and then executes the connected cruise control strategy [32].

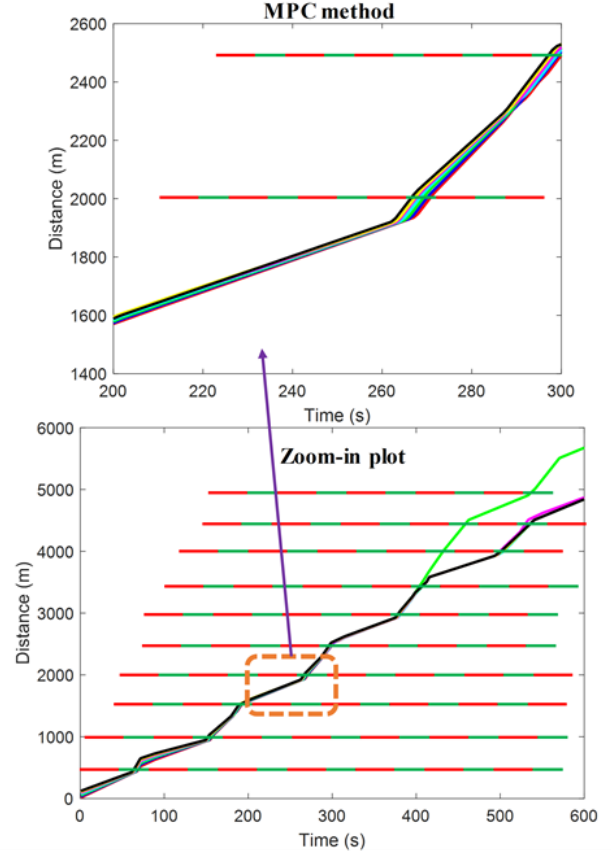

Fig. 6. Vehicle distance trajectories in MPC-based method for 10 minutes

Fig. 6 and Fig. 7 illustrate the vehicle trajectories (distance varies over time) in the MPC-based and baseline approaches for 10 minutes. The horizontal red bars depict the red lights duration time and the green bars point to the green lights' intervals. The vehicles can only go through the traffic signal during the green bars period. The zoom-in plots in these figures indicate that connected autonomous vehicles move together until the traffic red lights split them. In the baseline method, the vehicles split into two groups at 100 second, however, the vehicles in the MPC-based method split until 400 second. These results attribute to the limitation of road speed (the rest of vehicles cannot violate the limitation) and indicate that the MPC-based technique can significantly improve the system mobility.

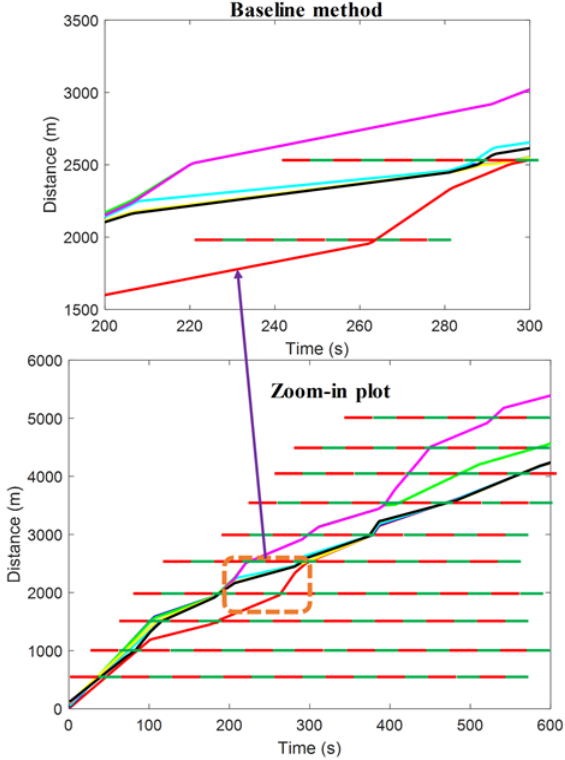

Fig. 7. Vehicle distance trajectories in baseline method for 10 minutes

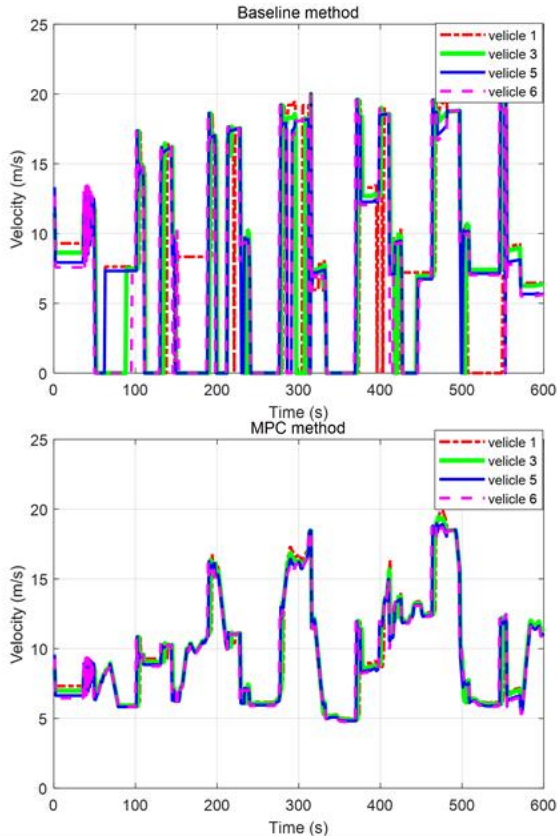

Fig. 8. Four vehicles' velocity profiles in two methods for 10 minutes

To avoid chaotic and unclear exhibition, four random vehicles from eight vehicles are selected to display their running speeds in the defined scenario. Fig. 8 describes the velocity profiles for these vehicles (their order: 1, 3, 5, 6, a total of 8 vehicles). It is apparent that the vehicle velocities in the MPC-based method always stay positive which indicates that all the vehicles would not stop at every traffic signal light. Oppositely, the connected vehicles controlled by the baseline method have to stop at all traffic signals. Thus, the traveling time of each vehicle (pass through the same driving scenario) could be reduced by 30% (shown in Table 3) in the proposed method. It also suggests that longitudinal speed decision-making control strategy presented in the higher-level controller could effectively avoid stopping at red signal lights so that the system mobility can be highly promoted.

*B. Evaluation of Integrated System*

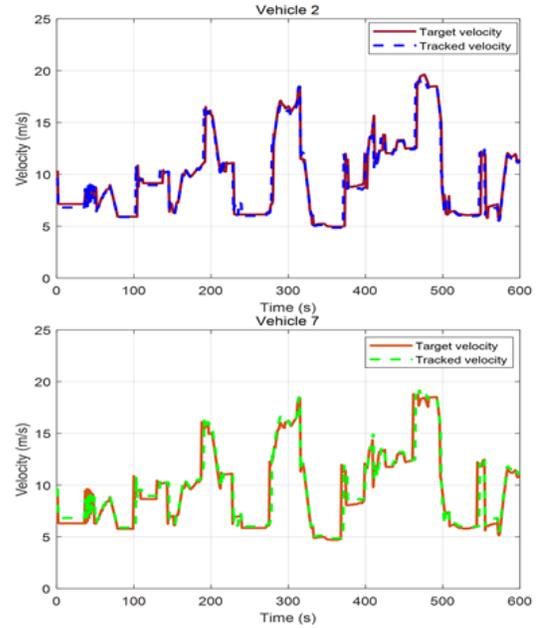

Fig. 9. Velocity tracking in vehicles 2 and 7 using the RL algorithm

In this section, the proposed whole system is compared with a benchmark system to assess its effectiveness and merits. In this benchmark (baseline) system, the higher-level controller is the same as our proposed MPC-based longitudinal speed decision-making control strategy, while the lower-level still utilizes the MPC method [28] instead of the RL algorithm. Similarly, to prevent tedious and chaotic demonstrations of relevant results, two random vehicles are chosen to compare their energy management strategies. The order of these two vehicles is 2 and 7 (totally eight vehicles).

The velocity tracking in both vehicles using the RL algorithm in the lower-level controller are shown in Fig. 9. It illustrates that the lower-level controller is able to track the optimal velocity trajectory provided by the higher-level controller almost perfectly. Under this situation, the proposed method can avoid stopping at red lights and guarantee improving energy efficiency simultaneously.

Fig. 10 shows the state variable SOC evolution for these two vehicles in different approaches. For the same vehicle, the derived velocity trajectories are different in these two methods.

Thus, the related power demand and control actions are different, which results in the disparate variation of SOC curves. Due to the charge sustenance constraint in the lower-level controller, the final SOC value is close to the initial value. This implies that the battery power used by the electrified vehicle, leading to a decrease of SOC, which is compensated by charging with the power from the engine and regenerative braking power. Hence, we can evaluate the different methods' control performance by comparing the power split and fuel consumption.

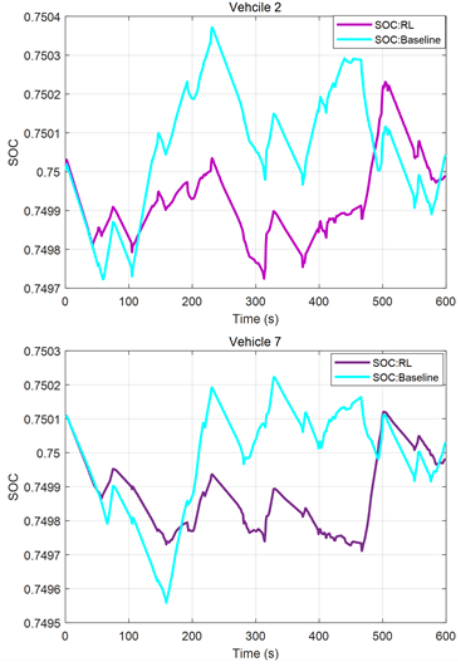

Fig. 10. SOC trajectories for vehicles 2 and 7 in two methods.

The power split controls in these two vehicles are described in Fig. 11. As Eqs. (9)-(12) stated, the battery and engine powers are finally decided by the throttle percentage. From Fig. 10, it is obvious that the power split controls in the proposed system are different from those of the benchmark system, which implies these two systems could achieve different control performance although the input speed profiles are the same. The working points of the engine in the different control strategies are given in Fig. 12. It is apparent that the engine working area under the proposed RL-based control locates in the lower fuel consumption region more frequently, compared to the MPC-based control. This implies that the RL method leads to higher energy efficiency than MPC technique. Hence, the proposed system could achieve lower fuel consumption compared with the baseline system.

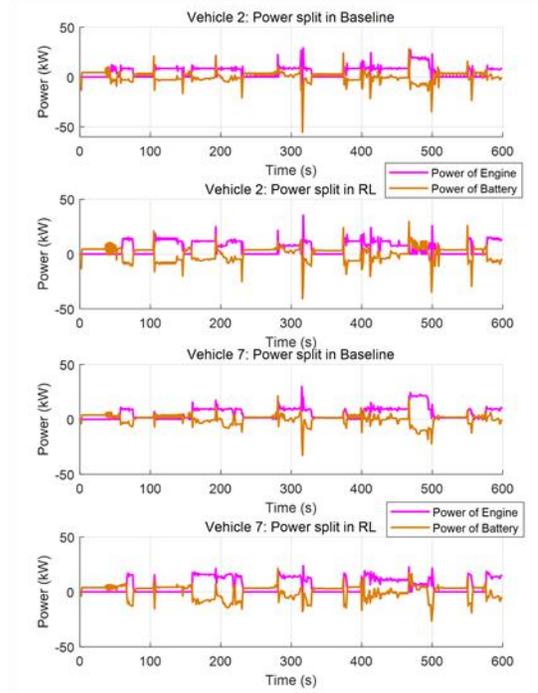

Fig. 11. Power split for vehicles 2 and 7 in two methods.

To quantify the performance in driving mobility and energy efficiency [33, 34], Table 3 shows the corresponding fuel consumption after SOC-correction and traveling time in these two systems for two selected vehicles (Vehicle 2 and 7). Note that, the corrected fuel consumption in the proposed system is lower than that of the baseline system for both vehicles. These results illustrate that our proposed energy efficiency control can achieve higher control performance than MPC-based control in the lower-level. For the same vehicle, it is obvious that the traveling time in the proposed system is shorter than that of benchmark system. It means the vehicle could reduce red light idling and improve the average running speed, and thus promote the driving mobility by using the proposed system. Thus, the proposed MPC and RL system can improve both driving mobility and energy efficiency.

Furthermore, to evaluate the possible application in re-al-world, Table 4 displays the computation time of the pro-posed system and benchmark system for the chosen two ve-hicles. It can be discerned that the proposed system is shorter than the baseline one and it is feasible to be applied online for automated HEV. The future work includes extending the MPC and RL-enabled system into multiple driving scenarios, and thus it can help the vehicle improve the mobility and fuel economy in real-world environments.

Table III
Fuel consumption and traveling time in different systems for two vehicles

| Order | Algorithms | Fuel consumption[a](g) | Saving(%) | Traveling time[b] (s) | Saving(%) |
|---|---|---|---|---|---|
| Vehicle 2 | Baseline | 137.1 | — | 685.9 | — |
|  | RL | 118.8 | 13.3 | 479.3 | 30.1 |
| Vehicle 7 | Baseline | 145.5 | — | 707.4 | — |
|  | RL | 126.4 | 13.1 | 496.5 | 29.8 |

[a] A 2.7 GHz microprocessor with 3.8 GB RAM was used.
[b] Traveling distance is 5000 m for each vehicle.

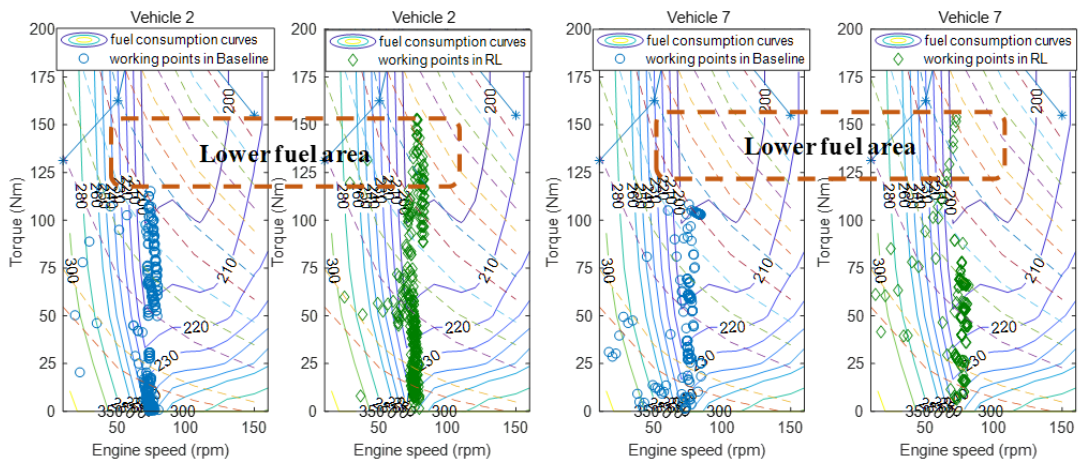

Fig. 12. Working points of engine for vehicles 2 and 7 in two methods.

Table IV
Operation time of compared systems for two chosen vehicles.

| Order | Systems | Computation time[a] (s) | Saving (%) |
|---|---|---|---|
| Vehicle 2 | Benchmark | 106.8 | — |
| | MPC+RL | 43.5 | 59.3 |
| Vehicle 7 | Benchmark | 112.7 | — |
| | MPC+RL | 48.9 | 56.6 |

[a] A 2.7 GHz microprocessor with 3.8 GB RAM was used.

Overall, our proposed longitudinal speed decision-making and energy efficiency system can not only improve the system mobility for a group of connected autonomous vehicles but also reduce the fuel consumption for each internal electrified vehicle. The reasons for these results are theoretically because the speed decision-making policy could efficiently reduce the red idling and the RL-based energy management strategy could reasonably allocate the output powers for multiple energy sources. As the computation time of the presented whole system is short and the RL algorithm could obtain model-free controls, the proposed hierarchical control architecture is promising to be applied in real-time.

## V. CONCLUSION

A hierarchical control architecture to address the longitudinal speed decision-making and energy efficiency control problem for connected autonomous electrified vehicles is presented in this paper. The higher-level controller can use the model predictive control approach to handle the signal phase and timing and state information of vehicles via vehicle and vehicle or infrastructure communication to improve the system mobility. The lower-level controller exploits the reinforcement learning algorithm to manage the optimal velocity profile from the higher-level to reduce the fuel consumption for each electrified vehicle.

It is verified in the tests that the proposed longitudinal speed decision-making and energy efficiency control strategy is effective and available. Connected autonomous vehicles can avoid stopping at red signal lights by the model predictive control method, and each can highly improve fuel economy via the reinforcement learning approach. This methodology can be applied to both full and semi-autonomous vehicles. The future research direction involves predicting the driver behavior using the higher-level controller and addressing complex decision-making tasks in the lower-level controller.

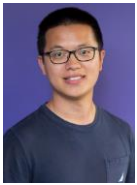

**Teng Liu** (M'2018) received the B.S. degree in mathematics from Beijing Institute of Technology, Beijing, China, 2011. He received his Ph.D. degree in the automotive engineering from Beijing Institute of Technology (BIT), Beijing, in 2017. His Ph.D. dissertation, under the supervision of Pro. Fengchun Sun, was entitled "Reinforcement learning-based energy management for hybrid electric vehicles." He worked as a research fellow in Vehicle Intelligence Pioneers Ltd for one year. Now, he is a member of IEEE VTS, IEEE ITS, IEEE IES, IEEE TEC and IEEE/CAA.Dr. Liu is now a postdoctoral fellow at Department of Mechanical and Mechatronics Engineering, University of Waterloo, Ontario N2L3G1, Canada. Dr. Liu has more than 8 years' research and working experience in renewable vehicle and connected autonomous vehicle. His current research focuses on reinforcement learning (RL)-based energy management in hybrid electric vehicles, RL-based decision making for autonomous vehicles, and CPSS-based parallel driving. He has published over 30 SCI papers and 10 conference papers in these areas. He received the Merit Student of Beijing in 2011, the Teli Xu Scholarship (Highest Honor) of Beijing Institute of Technology in 2015, "Top 10" in 2018 IEEE VTS Motor Vehicle Challenge and sole outstanding winner in 2018 ABB Intelligent Technology Competition. Dr. Liu is a workshop co-chair in 2018 IEEE Intelligent Vehicles Symposium (IV 2018) and has been reviewer in multiple SCI journals, selectively including IEEE Trans. Industrial Electronics, IEEE Trans. on Intelligent Vehicles, IEEE Trans. Intelligent Transportation Systems, IEEE Transactions on Systems, Man, and Cybernetics: Systems, IEEE Transactions on Industrial Informatics, Advances in Mechanical Engineering.

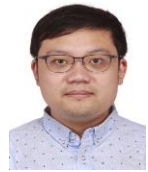

**Bo Wang** received B. S. degree in Mathematics from Beijing Institute of Technology, China in 2011. He received Ph. D. degree in Mathematics from Beijing Normal University, China, in 2016. He is currently an Assistant Professor at Beijing Institute of Technology, China. His research focuses on partial differential equations and their applications.

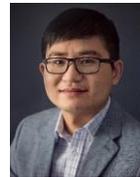

**Dongpu Cao** received the Ph.D. degree from Concordia University, Canada, in 2008. He is currently an Associate Professor and Director of Driver Cognition and Automated Driving (DC-Auto) Lab at University of Waterloo, Canada. His research focuses on vehicle dynamics and control, driver cognition, automated driving and parallel driving, where he has contributed more than 170 publications and 1 US patent. He received the ASME AVTT'2010 Best Paper Award and 2012 SAE Arch T. Colwell Merit Award. Dr. Cao serves as an Associate Editor for IEEE TRANSACTIONS ON VEHICULAR TECHNOLOGY, IEEE TRANSACTIONS ON INTELLIGENT TRANSPORTATION SYSTEMS, IEEE/ASME TRANSACTIONS ON MECHATRONICS, IEEE TRANSACTIONS ON INDUSTRIAL ELECTRONICS and ASME JOURNAL OF DYNAMIC SYSTEMS, MEASUREMENT AND CONTROL. He has been a Guest Editor for VEHICLE SYSTEM DYNAMICS, and IEEE TRANSACTIONS ON SMC: SYSTEMS. He has been serving on the SAE International Vehicle Dynamics Standards Committee and a few ASME, SAE, IEEE technical committees, and serves as Co-Chair of IEEE ITSS Technical Committee on Cooperative Driving.

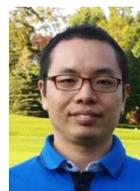

**Xiaolin Tang** received B.S. degree in Mechanics engineering and M.S. Degree in Vehicle Engineering from Chongqing University, China, in 2006 and 2009, respectively. He received the Ph.D. degree in Mechanical Engineering from Shanghai Jiao Tong University, China, in 2015. He is currently an Assistant Professor at the State Key Laboratory of Mechanical Transmissions and at the Department of Automotive Engineering, Chongqing University, Chongqing, China. His research focuses on Hybrid Electric Vehicles (HEVs), Vehicle Dynamics, Noise and Vibration, and Transmission Control.

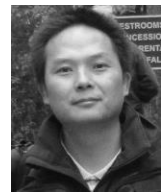

**Yalian Yang** received the Ph.D. degree in mechanical engineering from Chongqing University, Chongqing, China, in 2002. He is currently a Professor with the State Key Laboratory of Mechanical Transmissions, Chongqing University. His research interests include alternative energy drivelines, virtual reality driving simulator, and vehicle dynamic simulation and control.